# On Supporting Digital Journalism:
## Case Studies in Co-Designing Journalistic Tools

**Georgiana Ifrim**
School of Computer Science,
University College Dublin,
Dublin, Ireland
georgiana.ifrim@ucd.ie

**Derek Greene**
School of Computer Science,
University College Dublin,
Dublin, Ireland
derek.greene@ucd.ie

**Mark T. Keane**
School of Computer Science,
University College Dublin,
Dublin, Ireland
mark.keane@ucd.ie

**Claudia Orellana-Rodriguez**
Insight Centre for Data Analytics
University College Dublin,
Dublin, Ireland
claudia.orellana@insight-centre.org

**Bichen Shi**
Insight Centre for Data Analytics
University College Dublin,
Dublin, Ireland
bichen.shi@insight-centre.org

**Gevorg Poghosyan**
Insight Centre for Data Analytics
University College Dublin,
Dublin, Ireland
gevorg.poghosyan@insight-centre.org

**ABSTRACT**

Since 2013 researchers at University College Dublin in the Insight Centre for Data Analytics have been involved in a significant research programme in digital journalism, specifically targeting tools and social media guidelines to support the work of journalists. Most of this programme was undertaken in collaboration with *The Irish Times*. This collaboration involved identifying key problems currently faced by digital journalists, developing tools as solutions to these problems, and then iteratively co-designing these tools with feedback from journalists. This paper reports on our experiences and learnings from this research programme, with a view to informing similar efforts in the future.

## 1  INTRODUCTION

Beginning in 2013, several researchers (involving 5 Principal Investigators, 5 PhDs and 3 Post-Docs) in the School of Computer Science and the Insight Centre for Data Analytics at University College Dublin (UCD) worked on a substantial research program in Digital Journalism. This program built upon previous research on story tracking and news curation in social media [1]. In its development of tools for digital journalists, this program holds many lessons for the future of the field. Over a three-year period, a large collaborative project was conducted with a leading news provider in Ireland, *The Irish Times*, to identify emerging problems in digital journalism, focusing on how data analytics could be applied to address these problems. In this program, problem areas were jointly scoped by data analytics researchers and newsroom teams. These problems were then evaluated, technically and pragmatically, before being selected as key projects. Then, novel tools were developed to solve these problems. All of the projects developed novel digital-tools, that were subsequently iteratively-designed and evaluated in conjunction with journalists at *The Irish Times*. Many of the tools have won international data-science competitions and have been presented at major international conferences/workshops in AI and Machine Learning. The present paper sketches some of the successes of this research program while surfacing some of the learnings that have emerged of relevance to digital journalism.

## 2  A TALE OF THREE TOOLS & SOME GUIDELINES

At the start of the programme, we workshopped over 20 target problems defined by editorial and journalistic staff at *The Irish Times*. Following a technical and practical evaluation of these target areas (i.e., to determine which where feasible within the timelines of the programme) these 20 areas were reduced to 5 targets (n.b., there were other successful projects not reported here), described below.

*Hashtagger:* This tool was aimed at tracking news and Twitter feeds in real-time, to recommend emerging hashtags to news articles, thus enabling journalists to more effectively promote their content on Twitter. To deliver high-precision, real-time, hashtag recommendations, the tool employed learning-to-rank and automated feature-extraction methods (see [2], [3], [4]). It further employed scalable clustering and filtering techniques that won 1st prize in an international data challenge competition (see [2]). The recommendations and tracked feeds were delivered via a web application with several user-friendly features, such as sorting and saving articles, editing and tweeting hashtagged news headlines, along with the summarisation of hashtag usage trends [4].

*Headlines:* This tool was designed to provide feedback to editor-users on the "goodness" and SEO characteristics of headlines for news stories. The tool had an intuitive, web-based interface that took the headline-story as a feed and performed an analysis for the entities in the headline/story providing a measure of the goodness of different words using a large, corpus of previous news articles and frequency analyses of the terms in the target story [5].

*Topy:* This tool was developed to enable journalists to automatically assemble and curate story-specific webpages on topics of interest (e.g., the 2016 Irish General Election). A key novelty of the tool was its use of Twitter hashtags as *social indexes* in customised news-retrieval. This tool enabled the user to formulate queries that mix keywords and hashtags to refine the query (e.g., "UK #brexit"). The tool was a web-based application, allowing journalists to track updates for multiple queries/stories over time, for both short-lived news events (i.e., over weeks) and long-running stories (i.e., months to years; see [6], [7]).

*Tweeting Guidelines*: Following an analysis of the relative levels of attention that different news-related Twitter accounts attract in the Irish news eco-system, the programme quickly identified the importance of developing appropriate tweeting policies for newsrooms (i.e., when and how to tweet news content). Based on an extensive analysis of several large Twitter crawls covering accounts from Ireland and the United Kingdom, it was demonstrated that successful tweeting of news content varied consistently across different news categories, depending on time-of-day and time-of-week, and required a personal touch on the part of the tweeting journalistic account (i.e., corporate accounts were less successful). From this analysis, a set of guidelines were developed for journalists to tweet their news optimally (see [8]).

## 3   LESSONS LEARNED

There are many lessons that we have learned during this research programme that can only be sketched here. Overall the collaborative experience was synergistic and positive. The somewhat *ad hoc* structure of the group doing the co-design worked well in the problem-definition and early-design stages, but it was clear that more formal structures were required at later stages (especially, during user-evaluation). During evaluation, significant and dedicated time is required for potential end-users to assess a given tool; if personnel are not explicitly dedicated to this phase and time allotted to this task, then the evaluations are unlikely to be useful or informative. To be more positive, the problem specification and iterative-design stages generally worked well, once there was openness and regular communication from both sides. The most important learning during these stages was the importance of killing-off marginal ideas at an early point. That is, the computer scientists needed to kill off technically infeasible ideas, while journalists needed to kill off impractical ideas. Perhaps the largest unanticipated issue in tool development was the importance of performance characteristics. For instance, it was not enough for a tool to perform some complex task, it often also needed to be able to do that task within a restricted time-frame (i.e., given the time-critical nature of news production). As such, much of the later work in tool development concentrated on achieving speed-ups (i.e., much more D than R, in R&D). Finally, we also found that prior-supporting analysis of problem statements was of critical importance. Often newsrooms harbour anecdotal views on target issues (e.g., tweets could be sent at any time of the day), that are actually *not* supported by the data. So, a significant part of the research programme was the development of critical analyses to debunk/support anecdotal journalistic impressions in advance of tool development. Finally, the pace of change was always a significant issue; constantly changing requirements (on a timescale of weeks and months) persistently threatened to outpace the speed of the research efforts (on a timescale of years).

## ACKNOWLEDGMENTS
This work was supported by Science Foundation Ireland (SFI) under Grant Number SFI/12/RC/2289 and by funding from *The Irish Times*.